\documentclass[journal,comsoc]{IEEEtran}
\usepackage{algorithm,amsbsy,amsmath,amssymb,epsfig,bbm,mathrsfs,multirow,amsthm}
\usepackage{color,amsmath}
\usepackage{algpseudocode, tabularx}
\usepackage{graphicx,caption,subcaption,array,multirow,bm}
\usepackage{xcolor, tcolorbox}
\usepackage{makecell}
\usepackage{cite}
\usepackage{threeparttable}
\usepackage{diagbox}

\usepackage[labelformat=simple]{subcaption}

\DeclareMathAlphabet{\mathpzc}{OT1}{pzc}{m}{it}

\makeatletter
\newcommand{\multiline}[1]{%
	\begin{tabularx}{\dimexpr\linewidth-\ALG@thistlm}[t]{@{}X@{}}
		#1
	\end{tabularx}
}

\makeatother
\begin{document}
\title{A Hybrid Quantum-Classical Autoencoder Framework for End-to-End Communication Systems}

\author{Bolun Zhang, Gan Zheng, \IEEEmembership{Fellow,~IEEE}, Nguyen Van Huynh, \IEEEmembership{Member,~IEEE}
\thanks{Bolun Zhang and Gan Zheng are with the School of Engineering, University of Warwick, Coventry, CV4 7AL, UK. (Email: bolun.zhang@warwick.ac.uk, gan.zheng@warwick.ac.uk).

Nguyen Van Huynh is with the Department of Electrical Engineering and Electronics, University of Liverpool, Liverpool L69 3GJ, UK. (Email: huynh.nguyen@liverpool.ac.uk).
}}

\maketitle
\thispagestyle{empty}

\begin{abstract}
This paper investigates the application of quantum machine learning to End-to-End (E2E) communication systems in wireless fading scenarios. We introduce a novel hybrid quantum-classical autoencoder architecture that combines parameterized quantum circuits with classical deep neural networks (DNNs). Specifically, we propose a hybrid quantum-classical autoencoder (QAE) framework to optimize the E2E communication system. Our results demonstrate the feasibility of the proposed hybrid system, and reveal that it is the first work that can achieve comparable block error rate (BLER) performance to classical DNN-based and conventional channel coding schemes, while significantly reducing the number of trainable parameters. Additionally, the proposed QAE exhibits steady and superior BLER convergence over the classical autoencoder baseline.
\end{abstract}

\begin{IEEEkeywords}
End-to-End communication systems, quantum machine learning, parameterized quantum circuit, fading channel, and deep learning.
\end{IEEEkeywords}


\section{Introduction}
\label{Sec:intro}

\begin{figure}[!]
	\centering
	\begin{subfigure}[b]{0.49\textwidth}
		\centering
		\includegraphics[width=\textwidth]{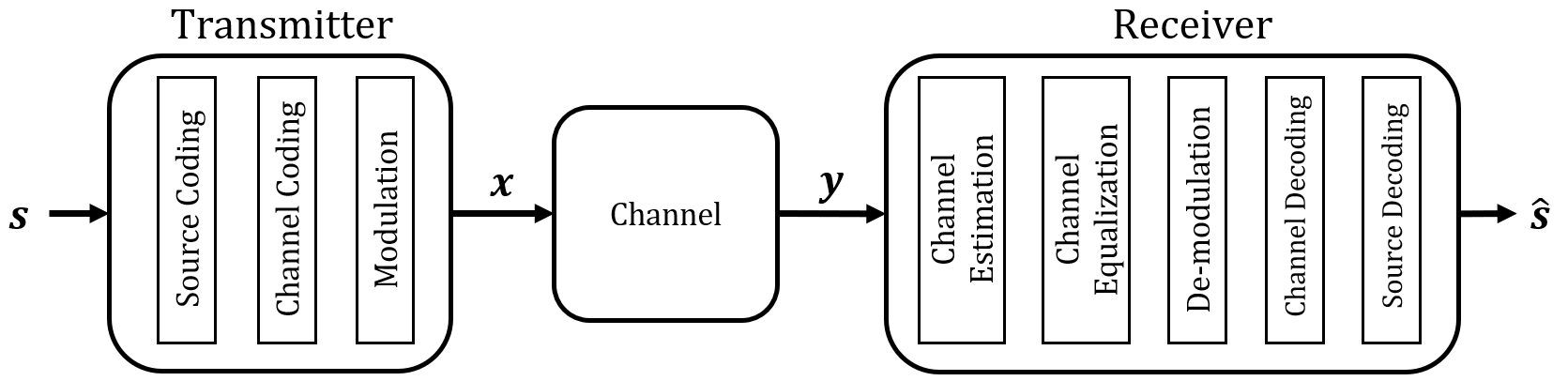}
		\caption{}
            \label{fig:arch_a}
	\end{subfigure}

	\begin{subfigure}[b]{0.49\textwidth}
		\centering
		\includegraphics[width=\textwidth]{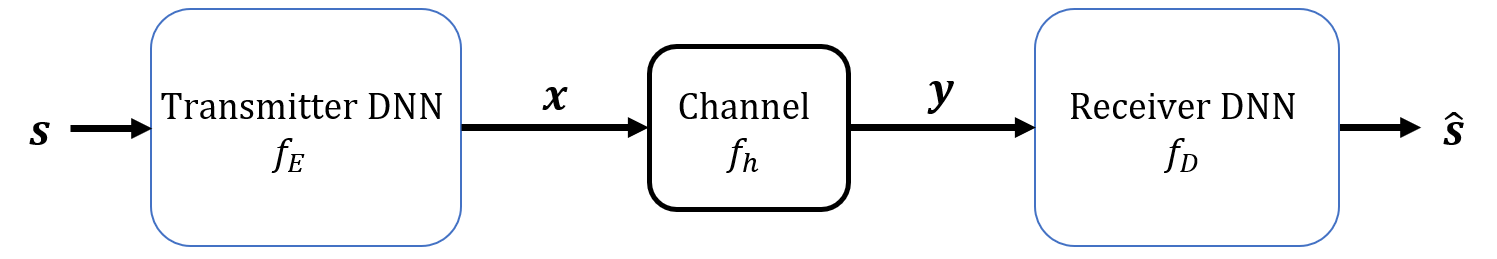}
		\caption{}
            \label{fig:arch_b}
	\end{subfigure}
	\caption{Architectures of (a) traditional communication systems and (b) Autoencoder-based E2E communication systems.}
	\label{fig:architecture}
\end{figure}

\IEEEPARstart{R}{\lowercase{ecently}}, deep learning (DL) has emerged as a promising approach for applications in wireless communication networks. Traditional communication systems rely on mathematically formulated signal processing blocks, which are often provably optimal for specific tasks, as shown in Fig. \ref{fig:arch_a}. However, the multi-block design makes it challenging to find an optimal and concrete transceiver configuration to learn the feature representation, especially in the real-world scenarios where the complex systems contain unknown effects that are difficult to model \cite{8psk}. Different from the conventional methods, DL-based approaches replace these rigid signal processing blocks with deep neural networks (DNNs) at both the transmitter and the receiver. This allows for joint optimization of transceivers without the need for explicitly designing complex mathematical models. By interpreting the communication system as an autoencoder, where the noisy channel is perceived as an intermediate layer connecting the transmitter and the receiver, the DL-based End-to-End (E2E) system can therefore be trained in a pure data-driven manner with a reconstruction loss function to jointly optimize both the transceivers without manually fine-tuning each signal processing block. 

The E2E paradigm is pioneered in \cite{intro_phy}, where the authors have proposed a foundational channel autoencoder architecture for the entire communication system, demonstrating near-optimal block error rate (BLER) performance compared to existing baseline modulation   and coding schemes. This E2E approach allows the system to learn effective solutions for channel impairments where the optimal strategy is unknown. It has proven to be effective in \cite{ofdm-autoencoder}. The use of one-hot embedding provides a simple yet effective framework for optimizing the E2E communication system in terms of BLER. However, the proposed autoencoder architecture primarily consists of multiple fully-connected (FC) layers, which causes the system's parameter size to grow exponentially with increasing block sizes, leading to increased computational complexity and reduced memory efficiency during training and inference.

Recently, quantum machine learning (QML) has gained significant attention in the field of wireless communications. The principles of quantum superposition and quantum entanglement offer a fundamental rethinking of traditional approaches, providing new insights to enhance the mainstream DNN-based architecture. Due to the unique properties of quantum mechanics, a quantum bit (qubit) can be represented as the superposition of both binary states, 0 and 1, simultaneously. This feature enables greater memory efficiency and a faster convergence rate compared to classical binary computing. In \cite{zhang2024hybrid}, the authors have proposed a novel hybrid quantum-classical architecture for downlink beamforming optimization that can achieve comparable performance as the classical DL-based scheme with significant parameter savings. In \cite{tabi2022hybrid}, the hybrid quantum-classical autoencoder for E2E communication systems is explored. However, the study only considers a single channel use, which is impractical in experimental settings. In addition, the transmitter of baseline lacks FC layers, relying solely on simple linear embedding, which does not adhere to the standard architecture in \cite{intro_phy}. Moreover, the error rate curves exhibit too much fluctuation, failing to show a consistent and interpretable trend. Given the above, we propose a hybrid quantum-classical autoencoder (QAE) framework under reasonable experimental settings with stable performance. Our key contributions are summarized as follows:

\begin{itemize}
    \item We develop a hybrid quantum-classical autoencoder framework that employs parallel quantum circuits at the transmitter. By utilizing the quantum superposition principle, the hybrid system alleviates the issue of exponentially growing parameters found in traditional classical autoencoder (AE) systems, achieving significant parameter savings without compromising BLER performance.

    \item We conduct extensive simulations to evaluate the BLER performance of the proposed hybrid architecture and demonstrate its advantages across various fading channels. The results show comparable BLER performance and a significant reduction in parameter size, with approximately 50\% fewer parameters compared to the classical autoencoder baseline \cite{intro_phy}.
    
    \item The proposed QAE demonstrates superior BLER convergence compared to the classical AE baseline. It also reveals that the parallel quantum circuit design can learn more effective mappings from one-hot vector to encoded signals, compared to the classical AE solution.
\end{itemize}


\section{System Model}
\label{sec:Problem_Formulation}

We consider an autoencoder-based E2E communication system, consisting of a transmitter, a channel, and a receiver, as depicted in Fig. \ref{fig:arch_b}. The transmitter and receiver are modeled as two feedforward DNNs, while the channel is represented as a non-trainable intermediate layer connecting both the transmitter and the receiver. In particular, the transmitter encodes the source message $\bm{s} \in \mathbb{M}$ into $n$ complex baseband symbols $\bm{x} \in \mathbb{C}^n$, subject to a fixed power constraint, i.e., $||\bm{x}||^2 \leq n$. In this context, $\mathbb{C}^n$ denotes the $n$-dimensional complex-valued representations, and $\mathbb{M}=\{0,1,...,M-1\}$ denotes the set of all possible messages. The receiver aims to decode the impaired received signal $\bm{y} \in \mathbb{C}^n$ and produce the estimates $\hat{\bm{s}}$ of the source message. Each source message $\bm{s}$ is encoded as an $M$-dimensional one-hot vector, where $k=\log_2(M)$ represents the number of bits per message. The communication rate is defined as $R = k/n$ in bits per channel use, and the notation $(n,k)$ refers to $k$ bits over $n$ channel uses. The channel layer is trained with a fixed noise power $\sigma^2$ per complex symbol. The received signal can be expressed as:
\begin{equation}
\bm{y}=h\bm{x}+\bm{w},
\label{eq:0}
\end{equation}
where $h$ represents the complex channel coefficient, $\bm{w} \sim \mathcal{CN}(0, \sigma^2\bm{I})$ denotes complex-valued additive Gaussian noise vector, and $\bm{I}$ refers to the identity matrix.

The output of the receiver is a probability distribution over all possible messages, with the index of the highest-probability element corresponding to the estimated message $\hat{\bm{s}}$. The process of the E2E communications can be illustrated as the sequence of three cascaded functions:
\begin{equation}
\hat{\bm{s}}=f_\mathrm{D}(f_\mathrm{h}(f_\mathrm{E}(\bm{s}; \bm{\theta}_\mathrm{E})); \bm{\theta}_\mathrm{D}),
\label{eq:1}
\end{equation}
where $f_\mathrm{E}$ represents the encoder function that maps the original source message $\bm{s}$ to the encoded signal, defined as $\bm{x}=f_\mathrm{E}(\bm{s}; \bm{\theta}_\mathrm{E})$, where $\bm{\theta}_\mathrm{E}$ denotes the trainable parameters of the transmitter neural network (NN). The function $f_\mathrm{h}$ describes the channel impairments, with $h$ representing the channel response, i.e., $\bm{y}=f_\mathrm{h}(\bm{x})$. Meanwhile, $f_\mathrm{D}$ is the decoder function tasked with recovering the received signal into the estimated message, defined as $\hat{\bm{s}}=f_\mathrm{D}(\bm{y}; \bm{\theta}_\mathrm{D})$, where $\bm{\theta}_\mathrm{D}$ refers to the trainable weights of the receiver NN. The objective of the E2E communication system is to jointly optimize the transmitter and receiver NNs by minimizing the E2E loss function, expressed as $L=\mathcal{L}(\bm{s}, \hat{\bm{s}})$. The loss function $\mathcal{L}(\bm{s}, \hat{\bm{s}})$ serves as the objective function that quantifies the distance between the original message $\bm{s}$ and the estimated message $\hat{\bm{s}}$, reflecting the accuracy of data recovery during transmission.

\section{Hybrid Quantum-Classical Autoencoder for End-to-End Communication Systems}
\label{sec:DDPG_E2E}

\subsection{Quantum Neural Networks}
QML is a promising emerging research area that combines the strengths of quantum computing (QC) and machine learning (ML) techniques. By leveraging quantum phenomena such as superposition and entanglement, QML can offer advantages in memory efficiency and convergence rates compared to classical ML methods. The advent of variational quantum circuits (VQC) has enabled the integration of QC with existing ML algorithms, and many studies have utilized VQC as a quantum neural network (QNN), such as in variational classifiers \cite{blance2021quantum}. A typical QNN, or VQC, consists of three key components: quantum embeddings, parameterized quantum circuit (PQC), and quantum measurements. In particular, quantum embedding transforms the classical inputs into quantum states, embedding the data in a high-dimensional Hilbert space. The choices of quantum embedding method depend on the properties of the classical data. In this context, we will employ amplitude embedding, which encodes a normalized classical data vector $\bm{x}_{d} \in \mathbb{R}^N$ with a length of $N$ into the amplitudes of $n$-qubit quantum state $|\psi\rangle$, where $\mathbb{R}^N$ denotes $N$-dimensional real-valued representations. This can be expressed as:
        
\begin{equation}
|\psi\rangle = \frac{1}{\sqrt{\sum_{i=1}^{N} |x_i|^2}}\sum_{i=1}^{N} x_i|i\rangle,
\label{eq:2}
\end{equation}
where $N = 2^{n}$, $x_i$ is the $i$-th element of $\bm{x}_{d}$, and $|i\rangle$ represents the $i$-th computational basis state. In the context of E2E communication systems, the source message $\bm{s}$, consisting of $k$ information bits, is encoded into an $M$-dimensional one-hot vector to enhance the classification of transmitted messages, where $M=2^k$. Thus, amplitude embedding aligns perfectly with the use of one-hot encoding, as it translates the $M$-dimensional one-hot vector into the quantum state for processing in the PQC. According to quantum mechanics \cite{nielsen2010quantum}, a quantum state $|\Psi\rangle$, in Dirac notation, represents the state of a closed quantum system, which is a unit vector (i.e., $\langle\psi|\psi\rangle=1$) in Hilbert space. For a system of $n$ qubits, the quantum state $|\Psi\rangle$ is the tensor product of the individual qubits:

\begin{equation}
|\Psi\rangle = |\psi_1\rangle \otimes |\psi_2\rangle \otimes \cdots \otimes |\psi_n\rangle =
\sum_{p=00\cdots0}^{11\cdots1} h_p |p\rangle,
\end{equation}
where $|\psi_i\rangle$ represents the $i$-th qubit, a basic quantum information unit. The superposition principle allows the quantum state to be represented as a linear combination of all possible $n$-qubit states, with $h_p$ as the complex amplitude for each basis state $|p\rangle$. The tensor product $\otimes$ combines the individual qubits into an entangled quantum state. Since $h_p$ takes $2^n$ complex values, $|\Psi\rangle$ can be viewed as a superposition of $2^n$ eigenstates, ranging from $|00\cdots0\rangle$ to $|11\cdots1\rangle$. Therefore, an $M$-dimensional one-hot vector can be represented by $k=\log_2(M)$ qubits using the quantum properties. 

The following phase after quantum embeddings is PQC which consists of parameterized rotation gates and entangling operators. The parameterized rotation gates enable adjustments for each qubit state through tunable rotation angles. The tunable rotation angles correspond to the trainable weights in the classical DNNs. Meanwhile, the entangling operators create quantum entanglement between qubits, enhancing the correlation between qubits and enabling complex representations of circuits. The outputs of the PQC are obtained by measuring qubits' states in the computational basis, typically the Pauli-Z basis. This process transforms the quantum states back into classical data representations.

\subsection{Hybrid Quantum-Classical Autoencoder Architecture}

\begin{figure}[!]
  \centering
  \includegraphics[scale=0.3]{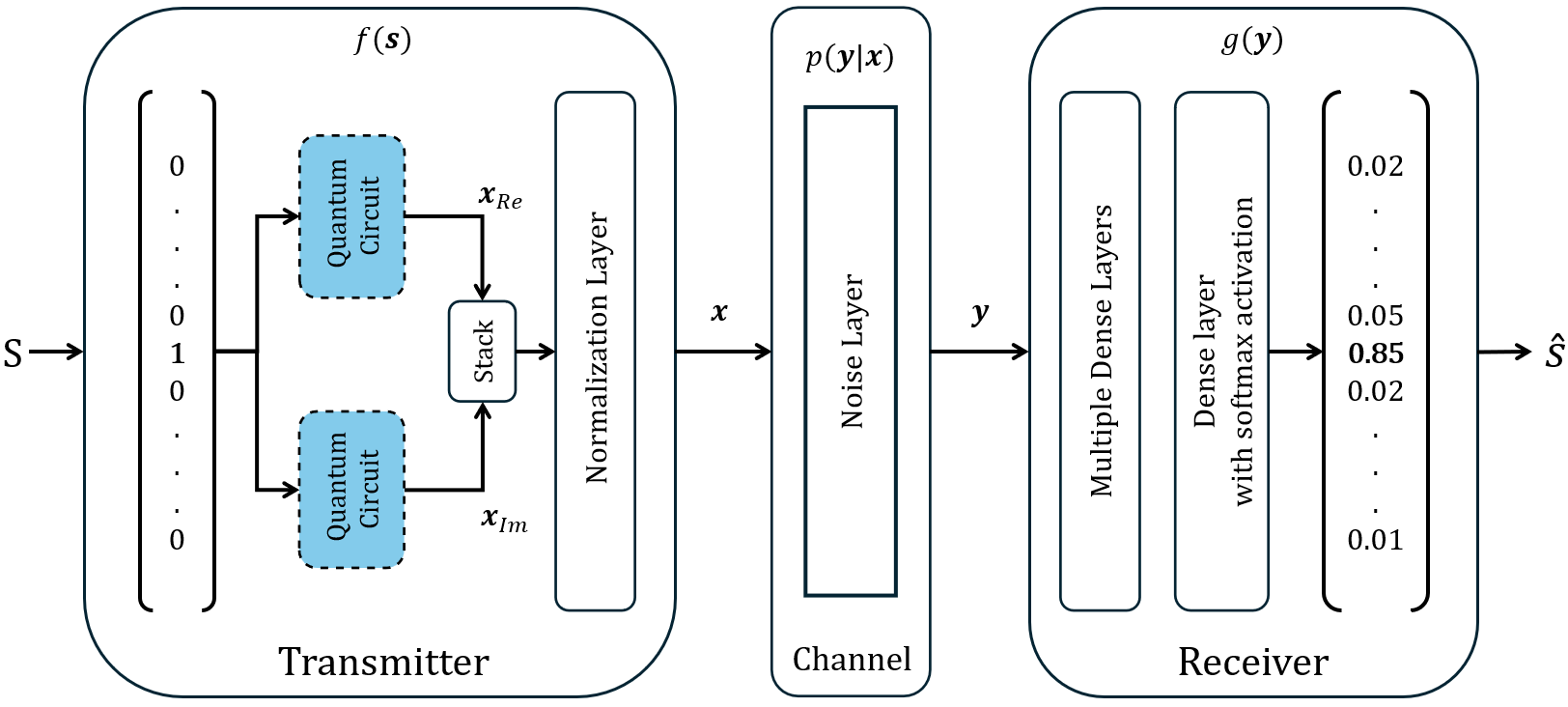}
  \caption{Overview of hybrid quantum-classical autoencoder.}
  \label{fig:qae_arch_parallel}
\end{figure}

In this paper, we propose a hybrid quantum-classical autoencoder architecture. Specifically, the transmitter consists of two parallel quantum circuits, each responsible for learning the real and imaginary components of the encoded signals. The receiver comprises multiple dense layers, with the final dense layer utilizing a softmax activation function to select the index with the highest probability for the estimated message $\hat{\bm{s}}$. The receiver architecture remains identical to the classical baseline. The schematic overview of the hybrid system is presented in Fig. \ref{fig:qae_arch_parallel}. In particular, for a $M$-dimensional input vector, the source message $\bm{s}$, containing $k$ information bits, is first converted into a one-hot vector. Two quantum circuits receive and process the one-hot vector simultaneously, with each circuit mapping the vector to the real and imaginary parts of the encoded signals, respectively. The two output states from the parallel quantum circuits, i.e., $\bm{x}_{\mathrm{Re}}$ and $\bm{x}_{\mathrm{Im}}$, are then stacked and normalized under a fixed power constraint, i.e., $||\bm{x}||^2 \leq n$. The received signals, after undergoing channel impairment, are processed by the FC-layer based receiver, which maps it to the probability distribution over $2^k=M$ possible messages. In this context, the number of channel uses $n$ corresponds to the number of qubits for the quantum circuit.

\begin{figure}[!]
  \centering
  \includegraphics[scale=0.32]{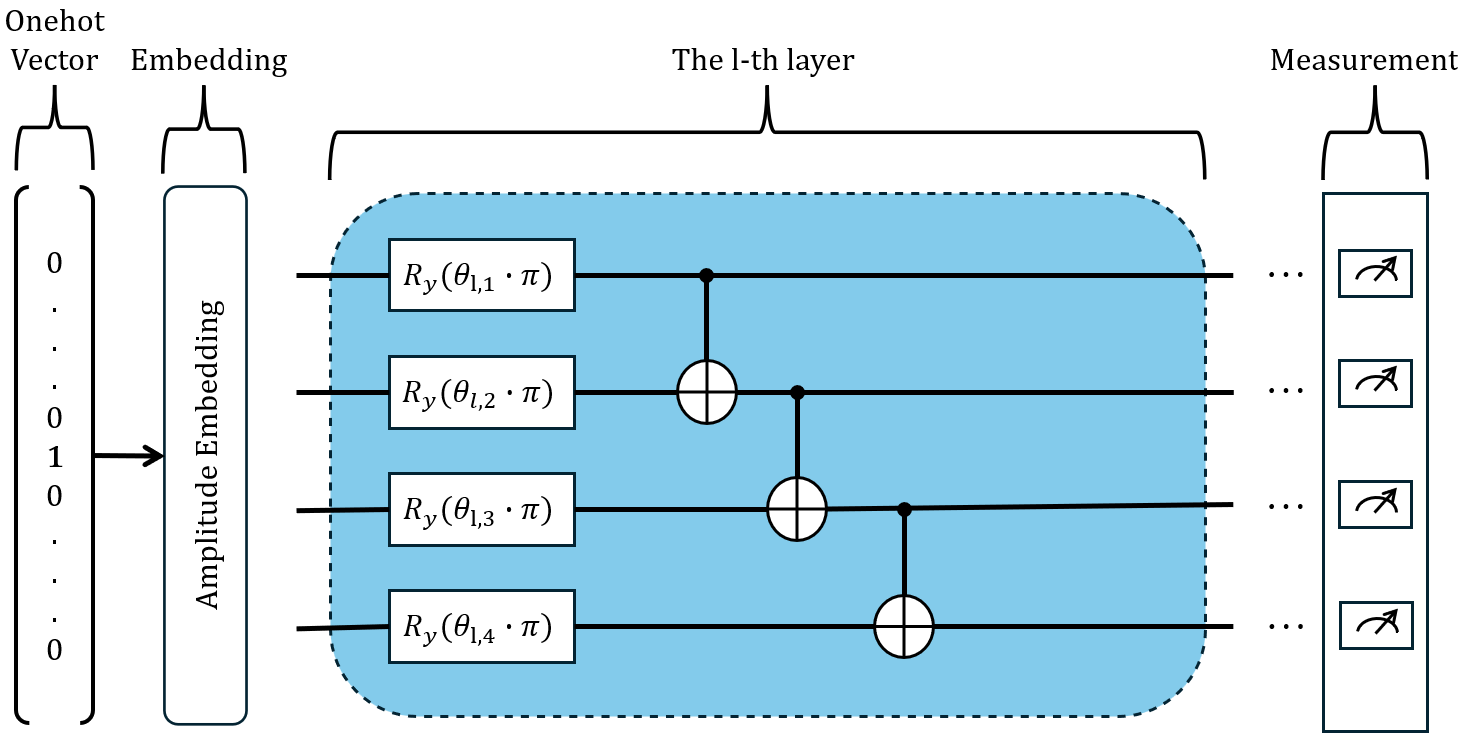}
  \caption{The $l$-th layer of quantum circuit with 4-qubit.}
  \label{fig:q_circuit}
\end{figure}

A detailed schematic of the QNN architecture is depicted in Fig. \ref{fig:q_circuit}. The one-hot vector is first converted into quantum states using amplitude embedding. Afterward, $L$ PQCs with identical structures are applied, with the $l$-th illustrated in the middle of Fig. \ref{fig:q_circuit}. It is assumed that each QNN layer utilizes $n$ qubits, where $n$ is the number of channel uses. For the configuration $(n,k)$, the quantum circuit can directly map the $2^k$-dimensional one-hot vector to $n$-dimensional space when $n=k$. For the case of $n>k$, the first $k$ qubits encode the one-hot vector via amplitude embedding, and the remaining $n-k$ qubits are initialized in the basis state $|0\rangle$. This also maps $2^k$-dimensional one-hot vector to the $n$-dimensional space. The encoded quantum state is first processed by $R_y$ gates, with the rotation angles multiplied by $\pi$, corresponding to the trainable weights in classical DNNs. Multiplying the angles by $\pi$ ensures the rotations are scaled appropriately within the natural range of qubit rotations, stabilizing the learning process by constraining the angle to a manageable range \cite{biamonte2017quantum}. After the quantum rotations, the qubits are entangled using $CNOT$ gates. In particular, the $i$-th qubit is the control qubit and the ($i+1$)-th qubit is the target qubit. The measurement layer is the expectation values in the $Z$-basis, which projects the quantum state onto the eigenstates of the Pauli Z matrix. The quantum state \( |\psi\rangle \) for an \( n \)-qubit system, where each qubit undergoes a rotation around the y-axis, can be expressed as:

\begin{equation}
|\psi\rangle = \left( \bigotimes_{i=1}^{n} R_y(\theta_{l,i}) \right) |e_j\rangle,
\end{equation}
where \(|e_j\rangle\) is the one-hot encoded quantum state, defined as:
\begin{equation}
|e_j\rangle = |j\rangle.
\end{equation}
Specifically, $|j\rangle$ denotes the $j$-th computational basis state corresponding to the index $j$ in the one-hot vector representation. The \( R_y \) rotations for the \( i \)-th qubit can be expressed as:

\begin{equation}
R_y(\theta_{l,i}) = \begin{pmatrix}
\cos\left(\frac{\theta_{l,i}}{2}\right) & -\sin\left(\frac{\theta_{l,i}}{2}\right) \\
\sin\left(\frac{\theta_{l,i}}{2}\right) & \cos\left(\frac{\theta_{l,i}}{2}\right)
\end{pmatrix},
\label{eq:Ry_matrix}
\end{equation}
where \( \theta_{l,i} \) denotes the rotation angle for the \( i \)-th qubit at the \( l \)-th layer of the quantum circuit.

\subsection{Parameter and Complexity Analysis}
The primary advantage of QML over the classical DL technique is that it requires significantly fewer trainable parameters, resulting in memory efficiency in training and the deployment on quantum computers. The parameter analysis for AE is presented in Table \ref{tab:tabel1}. At the transmitter, the classical AE consists of two FC layers. The input data dimension is $M$, corresponding to the one-hot vector. The first FC layer processes this input, maintaining the same output dimension of $M$ resulting in $(M+1)M$ parameters. The second FC layer then maps the $M$-dimensional output to a $2n$-dimensional space, adding $2n(M+1)$ parameters. In contrast, the proposed quantum circuit uses $L$ PQC layers, each requiring $n$ parameters for $n$-qubit $R_y$ rotation gates, resulting in a total of $2Ln$ parameters for two quantum circuits. This leads to a significant reduction in the number of parameters compared to the classical AE. In the setting defined by $(n,k)$, where $n$ indicates the number of channel uses and $k$ represents the block size, the parameter size of classical AE is given by $P_{AE} = 2^{2k+1}+(2n+1)2^{k+1}+2^k+2n$. The parameter size of the QAE is calculated as $P_{QAE} = 2^{2k} + (n + 1)2^{k+1} + 2Ln$. The proposed QAE scheme shows an exponential reduction in trainable parameters relative to the block size \(k\), i.e., $P_{AE}-P_{QAE}=2^{2k}+(2n+1)2^{k}+(2-2L)n$. Note that the receiver part remains unchanged for both AE and QAE, and the parameter size of the receiver is given by $P_{Rx} = 2^{2k} + (n + 1)2^{k+1}$. Table \ref{tab:tabel2} provides detailed parameter comparisons for three different configurations of \((n, k)\), specifically \((4,4)\), \((7,4)\), and \((8,8)\). In the proposed hybrid system, we utilize 3 quantum layers for each PQC, given that $L=3$. The results show that QAE achieves nearly a 50\% reduction in parameter size compared to the classical AE, highlighting the efficiency of the quantum approach. The parameter savings are particularly significant in the case of (8,8), with a reduction of nearly 70,000 parameters in total. The proposed hybrid quantum system is implemented using Pennylane \cite{bergholm2018pennylane}. This library provides a software interface to wrap the quantum circuit into a trainable neural network layer. This allows the encapsulated quantum circuit to be trained as an independent neural network and run on traditional binary computing devices. Our experiments indicate that quantum simulation using the PennyLane lightning plugin \cite{asadi2024} takes twice as long as classical DNNs. However, this gap could be reduced with more advanced quantum simulators, combined with data compression or optimized parallel execution.


\begin{table}[H]
    \centering
    \begin{tabular}{c|>{\centering\arraybackslash}p{2cm}    >{\centering\arraybackslash}p{2cm}}
    Layer & \textbf{Output Dimension} & \textbf{Parameters} \\ 
    \hline
    \hline
    Input & $M$ & \textit{---}\\
    Dense+ReLU & $M$ & $(M+1)M$\\
    Dense+Linear & $2n$ & $(M+1)2n$ \\
    Normalization & $2n$ & \textit{---} \\
    \hline
    Noise & $2n$ & \textit{---} \\
    \hline
    Dense+ReLU & $M$ & $(2n+1)M$\\
    Dense+Softmax & $M$ & $(M+1)M$\\
    \end{tabular}
    \caption{Dimension and parameters of classical autoencoder.}
    \label{tab:tabel1}
\end{table}

\vspace{-0.1cm}
\begin{table}[H]
    \centering
    \begin{tabular}{c|>{\centering\arraybackslash}p{2cm}    >{\centering\arraybackslash}p{2cm}>{\centering\arraybackslash}p{2cm}}
    \hline
    \multirow{2}{*}{\textbf{Scheme}} & \multicolumn{3}{c}{\textbf{Parameter Size}} \\
    \cline{2-4}
     & (4,4) & (7,4) & (8,8) \\
    \hline
    \hline
    AE & 824 & 1,022 & 140,048 \\
    \hline
    \textbf{QAE} & \textbf{440} & \textbf{554} & \textbf{70,192} \\
    \hline
    \end{tabular}
    \caption{Parameter size comparisons for schemes under settings of (4,4), (7,4), and (8,8).}
    \label{tab:tabel2}
\end{table}


\section{Simulation Results}
\label{sec:evaluation}
\begin{figure*}[!]
	\centering
	\begin{subfigure}[b]{0.329\textwidth}
		\centering
		\includegraphics[width=\textwidth]{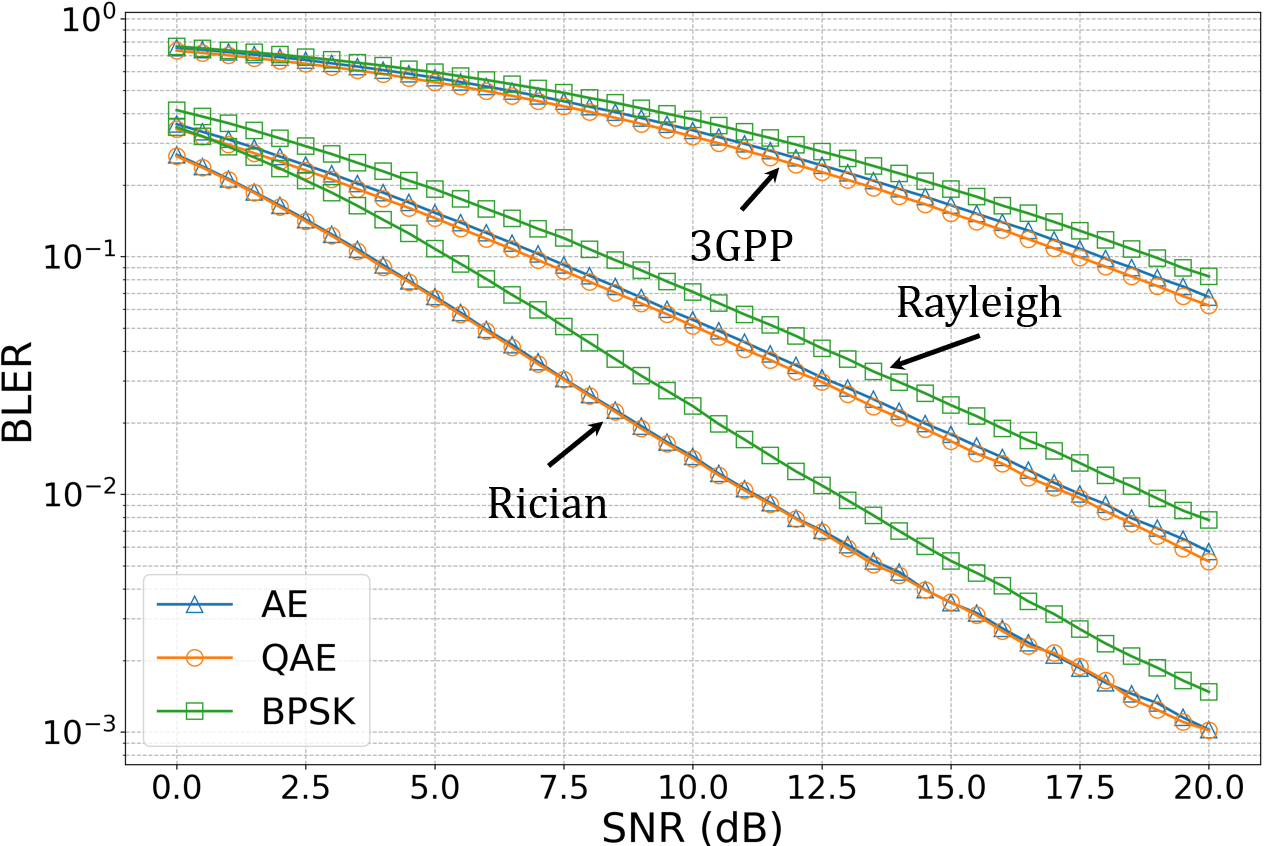}
		\caption{}
            \label{fig:bler1}
	\end{subfigure}
	\begin{subfigure}[b]{0.329\textwidth}
		\centering
		\includegraphics[width=\textwidth]{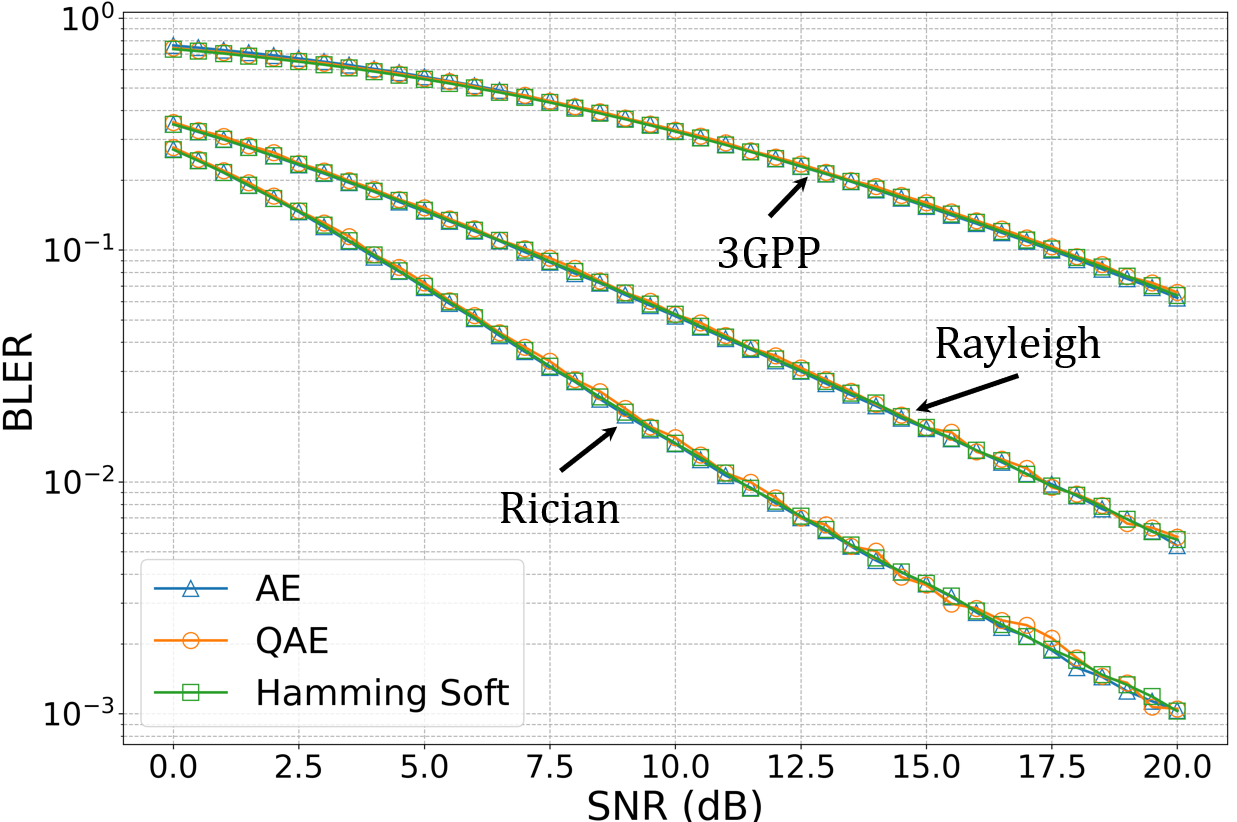}
		\caption{}
            \label{fig:bler2}
	\end{subfigure}
	\begin{subfigure}[b]{0.329\textwidth}
		\centering
		\includegraphics[width=\textwidth]{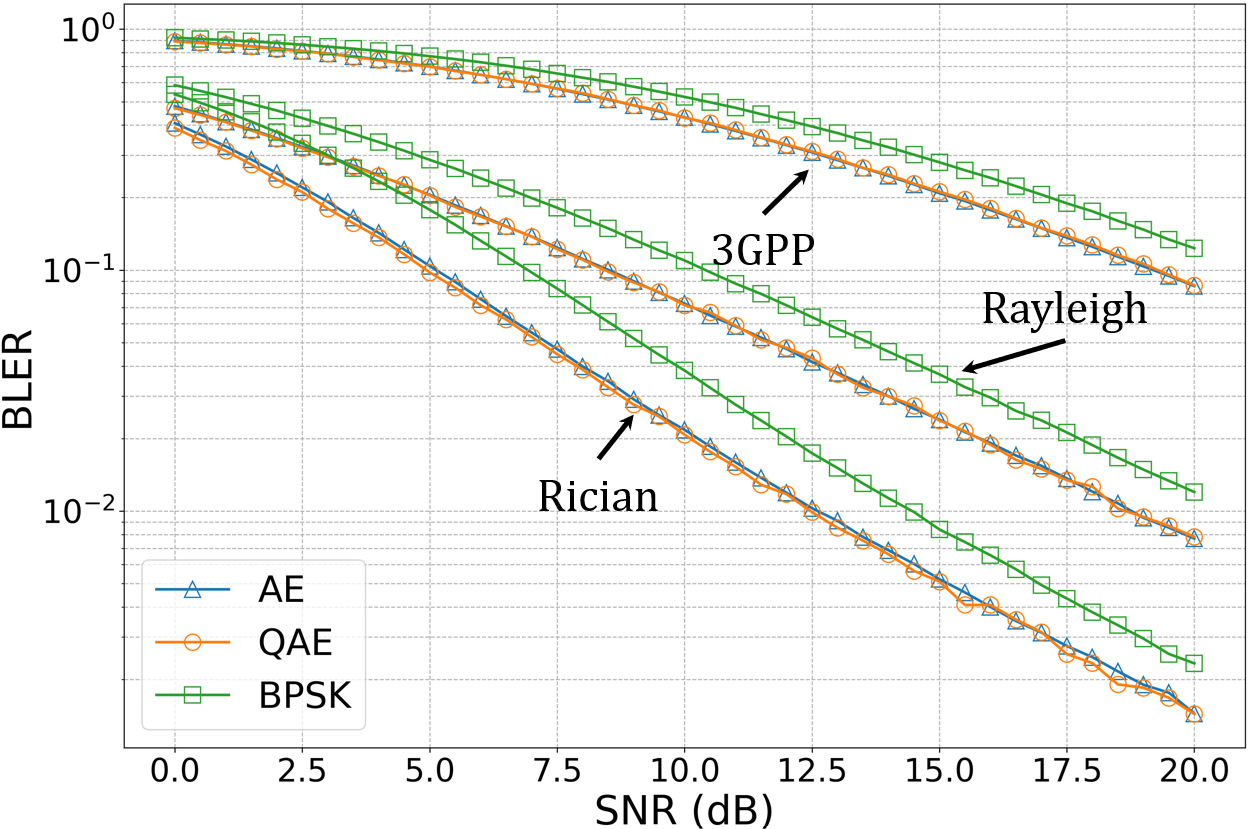}
		\caption{}
            \label{fig:bler3}
	\end{subfigure}
	\caption{BLER comparisons of (a) (4,4), (b) (7,4), and (c) (8,8) under Rayleigh, Rician, 3GPP block fading.}
	\label{fig:bler_joint}
\end{figure*}




\subsection{Experimental Settings}

In this section, we evaluate the experimental results of the proposed hybrid quantum-classical autoencoder framework. Classical AEs and conventional modulation or coding methods serve as baselines for comparison. The hyperparameters are aligned for both AE and QAE schemes to ensure fairness. Specifically, the batch size is set to 32, and the learning rate is set to 0.001. The Adam optimizer is used, along with the categorical cross-entropy loss function. Both systems are trained with a fixed SNR at 10 dB and evaluated across a range of SNRs from 0 to 20 dB. The systems are evaluated across three different $(n,k)$ settings: (4,4), (7,4), and (8,8). For the (4,4) and (8,8) scenarios, binary phase shift keying (BPSK) is employed as the baseline, while for the (7,4) configuration, Hamming code with soft decoding is used as the baseline. For the (7,4) configuration, the 7 qubits correspond to the number of channel uses, with the first 4 qubits used for the amplitude embedding to encode the one-hot vector to the quantum state, as the block size is 4. The remaining 3 qubits are initialized in basis state $|0\rangle$. Thus, PQC maps the $2^k$-dimensional one-hot vector into an $n$-dimensional quantum state. In configurations of (4,4) and (8,8), PQC directly maps $2^k$-dimensional one-hot vector into an $n$-dimensional quantum state, where $n=k$.

The BLER performance is assessed across different fading channels, including Rayleigh, Rician and 3GPP \cite{salo2005matlab}. For block fading channels, we assume the perfect channel state information and perform equalization, i.e., $\hat{x}=y/h$, where $\hat{x}$ is the estimated transmitted signal, $y$ is the received signal, and $h$ is the channel coefficients. The 3GPP model provides a practical channel for evaluating the proposed system, with the velocity of moving objects set to 30$km/h$, representing a moderate-speed vehicle. 

\subsection{Performance Evaluation}

\subsubsection{BLER vs SNR}
As shown in Fig. \ref{fig:bler1} for the (4,4) scenario, both the classical AE and the proposed QAE schemes outperform BPSK across all block fading channels. In particular, the proposed QAE system shows a slight BLER improvement over the AE, with a more noticeable performance gap in Rayleigh and 3GPP channels. In Fig. \ref{fig:bler2} for the (7,4) scenario, both the classical AE and QAE schemes exhibit nearly identical BLER performance to the Hamming code with soft decoding. This suggests that the proposed QAE scheme achieves comparable performance to the existing near-optimal channel coding baseline. Finally, in Fig. \ref{fig:bler3} for the (8,8) scenario, the proposed QAE system performs similarly to the AE scheme, with both surpassing the conventional BPSK. 

\subsubsection{BLER Convergence}
The QAE achieves a lower steady-state BLER after 20 epochs, and this trend remains consistent throughout the remaining range, demonstrating superior BLER convergence. The result further reveals that the parallel quantum circuits at the transmitter can learn a more effective mapping for encoding the one-hot vector to the transmitted signals than the classical DNN. This demonstrates the effectiveness of the proposed QAE framework in improving BLER performance while achieving significant parameter savings compared to the classical AE baseline.

\begin{figure}[!]
\centering
    \centering
    \includegraphics[width=0.47\textwidth]{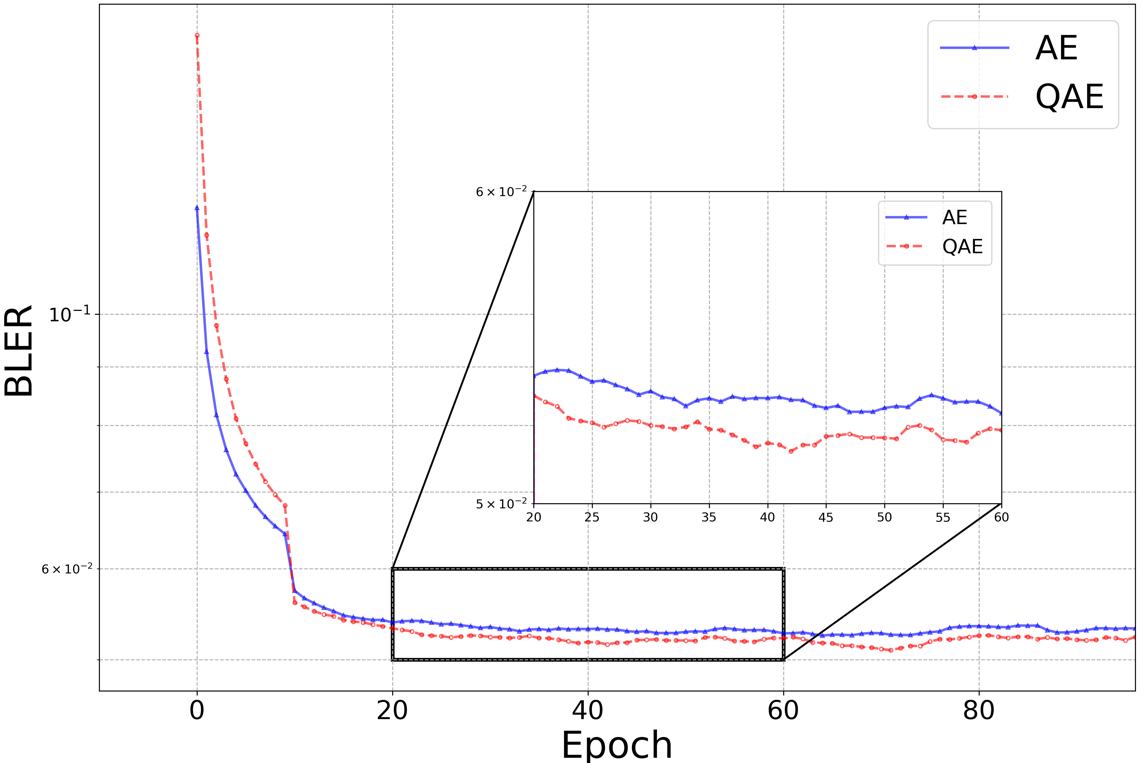}
\caption{BLER convergence (smoothed) comparisons between QAE and AE for Rayleigh block fading.}
    \label{fig:Convergence}
\end{figure}

\section{Conclusion}
\label{sec:conclusion}

In this paper, we introduce a hybrid quantum-classical autoencoder framework that exhibits advantages in both BLER performance and parameter savings compared to state-of-the-art solutions. Leveraging the quantum superposition principle, the quantum circuit encodes the large dimensional one-hot representation to a smaller dimensional quantum state using only a few qubits. Extensive simulations show that the proposed QAE system achieves BLER performance comparable to classical AE and conventional coding schemes across various fading scenarios, while significantly reducing the number of trainable parameters. Furthermore, the proposed QAE exhibits a superior BLER convergence advantage over the classical AE baseline throughout training. However, current quantum simulations on classical computers face challenges such as long execution time and exponentially increasing computational complexity due to the inherent demands of simulating quantum systems. In future work, we aim to leverage hardware-aware optimizations and more advanced quantum computing software to accelerate the computation of quantum simulators. Other future applications could investigate the use of quantum circuits in orthogonal frequency division multiplexing and multi-input multi-output systems. Additionally, integrating quantum convolutional neural networks into existing convolutional neural network-based solutions \cite{ye2019circular} could offer an alternative approach for handling large block sizes.

\begingroup
\footnotesize
\bibliographystyle{IEEEtran}
\bibliography{reference.bib}

\begin{thebibliography}{10}
\providecommand{\url}[1]{#1}
\csname url@samestyle\endcsname
\providecommand{\newblock}{\relax}
\providecommand{\bibinfo}[2]{#2}
\providecommand{\BIBentrySTDinterwordspacing}{\spaceskip=0pt\relax}
\providecommand{\BIBentryALTinterwordstretchfactor}{4}
\providecommand{\BIBentryALTinterwordspacing}{\spaceskip=\fontdimen2\font plus
\BIBentryALTinterwordstretchfactor\fontdimen3\font minus \fontdimen4\font\relax}
\providecommand{\BIBforeignlanguage}[2]{{%
\expandafter\ifx\csname l@#1\endcsname\relax
\typeout{** WARNING: IEEEtran.bst: No hyphenation pattern has been}%
\typeout{** loaded for the language `#1'. Using the pattern for}%
\typeout{** the default language instead.}%
\else
\language=\csname l@#1\endcsname
\fi
#2}}
\providecommand{\BIBdecl}{\relax}
\BIBdecl

\bibitem{8psk}
E.~Zehavi, ``8-{PSK} trellis codes for a rayleigh channel,'' \emph{IEEE Transactions on Communications}, vol.~40, no.~5, pp. 873--884, May. 1992.

\bibitem{intro_phy}
T.~O’shea and J.~Hoydis, ``An introduction to deep learning for the physical layer,'' \emph{IEEE Transactions on Cognitive Communications and Networking}, vol.~3, no.~4, pp. 563--575, Oct. 2017.

\bibitem{ofdm-autoencoder}
A.~Felix, S.~Cammerer, S.~D{\"o}rner, J.~Hoydis, and S.~Ten~Brink, ``Ofdm-autoencoder for end-to-end learning of communications systems,'' in \emph{2018 IEEE 19th International Workshop on Signal Processing Advances in Wireless Communications (SPAWC)}.\hskip 1em plus 0.5em minus 0.4em\relax IEEE, Jun. 2018, pp. 1--5.

\bibitem{zhang2024hybrid}
J.~Zhang, G.~Zheng, T.~Koike-Akino, K.-K. Wong, and F.~Burton, ``Hybrid quantum-classical neural networks for downlink beamforming optimization,'' \emph{IEEE Transactions on Wireless Communications}, Aug. 2024, \url{https://arxiv.org/abs/2408.04747}.

\bibitem{tabi2022hybrid}
Z.~Tabi, B.~Bak{\'o}, D.~T. Nagy, P.~Vaderna, Z.~Kallus, P.~H{\'a}ga, and Z.~Zimbor{\'a}s, ``Hybrid quantum-classical autoencoders for end-to-end radio communication,'' in \emph{2022 IEEE/ACM 7th Symposium on Edge Computing (SEC)}.\hskip 1em plus 0.5em minus 0.4em\relax IEEE, 2022, pp. 468--473.

\bibitem{blance2021quantum}
A.~Blance and M.~Spannowsky, ``Quantum machine learning for particle physics using a variational quantum classifier,'' \emph{Journal of High Energy Physics}, vol. 2021, no.~2, pp. 1--20, Feb. 2021.

\bibitem{nielsen2010quantum}
M.~A. Nielsen and I.~L. Chuang, \emph{Quantum Computation and Quantum Information}.\hskip 1em plus 0.5em minus 0.4em\relax Cambridge, UK: Cambridge University Press, Jan. 2010.

\bibitem{biamonte2017quantum}
J.~Biamonte, P.~Wittek, N.~Pancotti, P.~Rebentrost, N.~Wiebe, and S.~Lloyd, ``Quantum machine learning,'' \emph{Nature}, vol. 549, no. 7671, pp. 195--202, Aug. 2017.

\bibitem{bergholm2018pennylane}
V.~Bergholm, J.~Izaac, M.~Schuld, C.~Gogolin, S.~Ahmed, V.~Ajith, M.~S. Alam, G.~Alonso-Linaje, B.~AkashNarayanan, A.~Asadi \emph{et~al.}, ``Pennylane: Automatic differentiation of hybrid quantum-classical computations,'' \emph{arXiv preprint arXiv:1811.04968}, Nov. 2018.

\bibitem{asadi2024}
\BIBentryALTinterwordspacing
A.~Asadi, A.~Dusko, C.-Y. Park, V.~Michaud-Rioux, I.~Schoch, S.~Shu, T.~Vincent, and L.~J. O'Riordan, ``{Hybrid quantum programming with PennyLane Lightning on HPC platforms},'' 2024. [Online]. Available: \url{https://arxiv.org/abs/2403.02512}
\BIBentrySTDinterwordspacing

\bibitem{salo2005matlab}
J.~Salo, G.~Del~Galdo, J.~Salmi, P.~Ky{\"o}sti, M.~Milojevic, D.~Laselva, and C.~Schneider, ``Matlab implementation of the 3gpp spatial channel model (3gpp tr 25.996),'' \emph{on-line, Jan}, 2005.

\bibitem{ye2019circular}
H.~Ye, L.~Liang, and G.~Y. Li, ``Circular convolutional auto-encoder for channel coding,'' in \emph{2019 IEEE 20th International Workshop on Signal Processing Advances in Wireless Communications (SPAWC)}.\hskip 1em plus 0.5em minus 0.4em\relax IEEE, Jul. 2019, pp. 1--5.

\end{thebibliography}
\endgroup

\end{document}